\documentclass[9pt,twocolumn,twoside]{osajnl}

\journal{ol} 

\setboolean{shortarticle}{true}

\usepackage{siunitx}
\usepackage{color}
\usepackage{soul}

\usepackage{stfloats}

\def\eff{\text{eff}} 

\title{Probing the dispersive properties of optical fibers with an array of femtosecond-written fiber Bragg gratings}

\author[1,*]{Tommy Boilard}
\author[1]{Réal Vallée}
\author[1]{Martin Bernier}

\affil[1]{Center for Optics, Photonics, and Lasers (COPL), Université Laval, Québec (QC), Canada, G1V 0A6}

\affil[*]{Corresponding author: tommy.boilard.1@ulaval.ca}

\doi{}

\begin{abstract}
We propose an efficient method to determine the effective refractive index of step-index optical fibers from the visible to the mid-IR and thus allowing to infer their dispersive properties over a broad spectral range. The validity of the method, based on the writing of an array of fiber Bragg gratings (FBGs) with known periods using the fs scanning phase mask technique, is first confirmed with a standard silica fiber, then applied to various fluoride glass fibers to determine their effective refractive index and dispersion over more than three octaves, i.e. from 550 to \SI{4800}{\nano\meter}. 
\end{abstract}

\setboolean{displaycopyright}{false}

\begin{document}

\maketitle

There has been an increased interest in the last decade toward the development of pulsed fiber sources operating in the mid-infrared (mid-IR). Numerous applications are indeed calling for such bright and powerful laser sources with excellent beam quality and covering the so-called molecular fingerprint region. In molecular spectroscopy and environmental sensing, for example, several atmospheric pollutants including carbon oxides (CO, CO$_2$), hydrocarbons such as methane (CH$_4$) and nitrogen oxides (NO$_\text{X}$), possess absorption lines in the 3-\SI{5}{\micro\meter} spectral region that can be 10 to 100 times stronger than in the near-infrared (near-IR)~\cite{Gauthier2018}. This spectral region also happens to be an atmospheric transparency window, such that remote sensing of these molecules over long distance is feasible \cite{Tittel2007}.

Mid-IR ultrafast fiber sources based on direct light generation within femtosecond fiber laser~\cite{Duval2015, Hu2015, Huang2020, Bawden2021}, or based on soliton self-frequency shift (SSFS)~\cite{Tang2016, Duval2016} and supercontinuum generation (SC)~\cite{Xia2006, Domachuk2008, Petersen2014, Gauthier2016, Hudson2017} closely rely on the nonlinear and dispersive properties of the optical fibers to operate. Knowing, and eventually being able to precisely tailor, the shape of a dispersion curve over a broad spectral range extending up to the limit of transparency of the fiber material, is especially crucial for the optimization of such mid-IR sources. Now, although the physical properties, especially the refractive index, of silica fibers are now rather well known, those of fiber materials with extended transmission in the mid-IR, e.g. fluoride (including fluorozirconate (ZrF$_4$) and fluoroindate (InF$_3$)), tellurite and chalcogenide glasses are far less documented.

Actually, the refractive indices of mid-IR fibers are either theoretically inferred from their glass compositions~\cite{Zhang1994}, or directly measured from the fibers~\cite{Salem2015} or from bulk materials with a prism spectrometer, the validity of the latter being uncertain due to the decrease of the refractive index of the fiber upon drawing~\cite{Nakai1990}. Other means exist for determining directly the dispersion of optical fibers, such as by inferring it from the Kelly side lobes associated with ultrafast pulses propagating inside a ring cavity~\cite{Duval2015} or by using complex or dedicated interferometric setups~\cite{Ciacka2018, Klimentov2012}.

Another method that has been applied to highly sensitive silica-on-silicon waveguides is to write an array of weak FBGs  to infer their dispersion from the measurement of their effective refractive index from \SI{800}{\nano\meter} to \SI{1600}{\nano\meter}~\cite{Rogers2012}. In silica optical fibers, this method has only been limited to a mere \SI{50}{\nano\meter} spectral bandwidth~\cite{Cruz2021}, and its demonstration on a larger scale or in non-photosensitive fiber has yet to be made.

In this Letter, by writing an array of FBGs with precisely known periods using the femtosecond scanning phase mask technique, we accurately determine the dispersive properties of non-photosensitive specialty optical fibers over a bandwidth extending over more than three octaves, i.e. from the visible (\SI{550}{\nano\meter}) to the mid-IR (\SI{4800}{\nano\meter}). The approach consists in measuring the first, second and third diffraction orders of FBGs to retrieve the effective refractive index ($n_\eff$) spectral distribution of the fiber's fundamental mode, that is finally used to infer the dispersion of the fiber. Our results are first validated in a standard silica fiber (SMF-28, Corning), and then extended from the visible to the mid-IR in an erbium-doped fluorozirconate optical fiber and three undoped fluoroindate optical fibers. Finally, we propose to extend this method to characterize the dispersion of other exotic fibers and to precisely and fully characterize other optical and physical parameters of such fibers, namely their cutoff wavelength, their numerical aperture and even their core diameter by measuring the Bragg wavelength associated with higher order modes propagating in the fiber.

\begin{table*}[htbp]
\centering
\caption{\bf Fibers used in the experiment}
\begin{tabular}{cccccc}
\hline
Fiber name & Manufacturer & Host material$^\text{a}$ & Dimensions (\SI{}{\micro\meter}) & NA (-) & $\lambda_C$ (\SI{}{\micro\meter}) \\
\hline
SMF-28 & Corning & SiO$_2$ & 8.2/125 & 0.14 & 1.310 \\
ZFG DC [2.5] (ErF3 70000) 15/240*260/290 & Le Verre Fluoré & ZrF$_4$ & 15/240x260 & 0.125 & 2.5 \\
IFG SM [2.95] 7.5/125 & Le Verre Fluoré & InF$_3$ & 7.5/125 & 0.30 & 2.95 \\
IFG SM [3.3] 8.5/125 & Le Verre Fluoré & InF$_3$ & 8.5/125 & 0.30 & 3.3 \\
IFG SM [3.7] 9.5/125 & Le Verre Fluoré & InF$_3$ & 9.5/125 & 0.30 & 3.7 \\
\hline
\multicolumn{6}{l}{$^\text{a}$ \footnotesize The labels ZrF$_4$ and InF$_3$ refer to the main constituent of the corresponding multicomponent glasses.}\\
\end{tabular}
  \label{t:listeFibre}
\end{table*}

The fs scanning phase mask writing setup used for this experiment has been used extensively in the past decade to write strong and low losses FBGs through the protective coating of silica, chalcogenide and fluoride glass fibers~\cite{Bernier2014, Bernier2007, Bernier2012} and was further optimized to efficiently write distributed arrays of FBGs for sensing applications~\cite{Habel2017, Boilard2020}. In brief, a Ti:Sapphire amplifier (Astrella, Coherent) is used to generate \SI{30}{\femto\second} pulses at \SI{800}{\nano\meter} with a repetition rate of \SI{1}{\kilo\hertz}. The pulses are tightly focalized with an acylindrical lens ($f = \SI{8}{\milli\meter}$), through a phase mask, inside the core of an optical fiber. The period of the FBG is half of the period of the phase mask, i.e. $\Lambda = \Lambda_{\text{PM}/2}$. Due to the transparency of acrylate coatings at \SI{800}{\nano\meter}, FBGs are easily written through it, which is an important feature for writing in exotic, and sometimes brittle, optical fibers. To increase the overlap between the FBG and the mode inside the fiber, the lens is mounted on piezoelectric actuators and is scanned over the core of the fiber, thereby increasing the FBG's reflectivity. Finally, the lens and the beam can be moved along the length of the fiber with a high precision translation stage to control its length. 

For this experiment, we designed a custom phase mask that was fabricated by e-beam lithography with twenty-four \SI{5}{\milli\meter}-long uniform gratings with periods evenly distributed from \SI{1000.0}{\nano\meter} to \SI{3300.0}{\nano\meter} along its length. Instead of standard uniform phase masks, the benefit of this phase mask is that multiple FBGs can be written inside a single strand of optical fiber without the need for changing and aligning the phase masks each time. Each period of the phase mask is selected by the position of the lens (i.e. the translation stage), and the FBGs are written one at a time with a multi-pass procedure to reach the targeted reflectivity. This procedure was adopted, instead of a single scan over the entire length of the phase mask at once, because a deterioration of the spectral quality of the FBGs was observed when the laser beam overlapped two adjacent periods of the phase mask due to interference~\cite{Goebel2020}. In addition, it turned out that the first period of the phase mask was unusable, the main hypothesis being that its period was too close to the laser writing wavelength, increasing dramatically the phase mask zeroth order that is detrimental for FBG writing. Thus, in practice, only 23 periods were used for the experiment.

A basic requirement of our method is that the effective refractive index needs to be precisely determined over a broad spectral range, i.e. from the visible (550 nm) to the mid-IR (4800 nm). To that purpose, four different optical spectrum analyzers (OSA) from Yokogawa (AQ6373B, AQ6370C, AQ6376 and AQ6377) and three SC sources with different spectral coverage (400-\SI{2400}{\nano\meter}, 2000-\SI{3900}{\nano\meter} and 2800-\SI{5500}{\nano\meter}) were used. The visible OSA was calibrated by measuring the central wavelength of a He-Ne laser, while the others were calibrated by correcting their spectral response according to HITRAN database at \SI{1.3}{\micro\meter}, \SI{1.9}{\micro\meter} and \SI{2.7}{\micro\meter} with absorption lines of water vapor, plus the CO$_2$ absorption lines around \SI{4.2}{\micro\meter}.

It is also important to recall that two factors must be considered in inferring the value of $n_\eff$ from the measured Bragg wavelength. First, the intrinsic dc variation of the effective refractive index $\overline{\delta n}_\eff$ (related to the strength and visibility of the grating~\cite{Erdogan1997}) is resulting in a shift of the position of the nominal Bragg wavelength $\lambda_0$ according to $\lambda_0 = 2 \left( n_\eff + \overline{\delta n}_\eff \right) \Lambda$. Secondly, in order to hold the fiber in a straight position in front of the phase mask, a strain $\epsilon$ must be applied to it which is also resulting in a shift of $\lambda_0$. In fact, during inscription, the period of the written FBG always remains equal to half that of the phase mask and is independent of the strain. However, the effective refractive index decreases due to the strain-optic coefficient of the fiber, $P_e$, leading to a measured Bragg wavelength during inscription that is lower than $\lambda_0$, namely $\lambda_W = \left( 1 - P_e \epsilon \right) \lambda_0$. When the strain is released after inscription, the effective refractive index gets back to its original value, while the period of the written FBG contracts by an amount proportional to the initial strain applied, leading to a final Bragg wavelength that is $\lambda_F = (1 - \epsilon) \lambda_0$. Since measuring one of these wavelengths (writing or final) leads to an underestimation of $n_\eff$, this phenomena was effectively compensated for by monitoring both wavelengths to determine $\lambda_0$. Regarding the shift in $\lambda_0$ due to $\overline{\delta n}_\eff$, it is taken into account by simulating the spectrum of each individual FBG according to the equations found in~\cite{Erdogan1997}, allowing for a precise determination of the parameters of each FBG. Note that the value of $\overline{\delta n}_\eff$ and its corresponding detrimental effect for precise effective refractive index determination can be greatly reduced by inscribing weak FBGs. Accordingly, the writing parameters were chosen such that the transmission dip of the \SI{4}{\milli\meter}-long FBGs never exceed \SI{-2}{\decibel}, corresponding to a maximum $\overline{\delta n}_\eff$ of \SI{2.4e-4}{}.

\begin{figure}[!ht]
    \centering
    \includegraphics{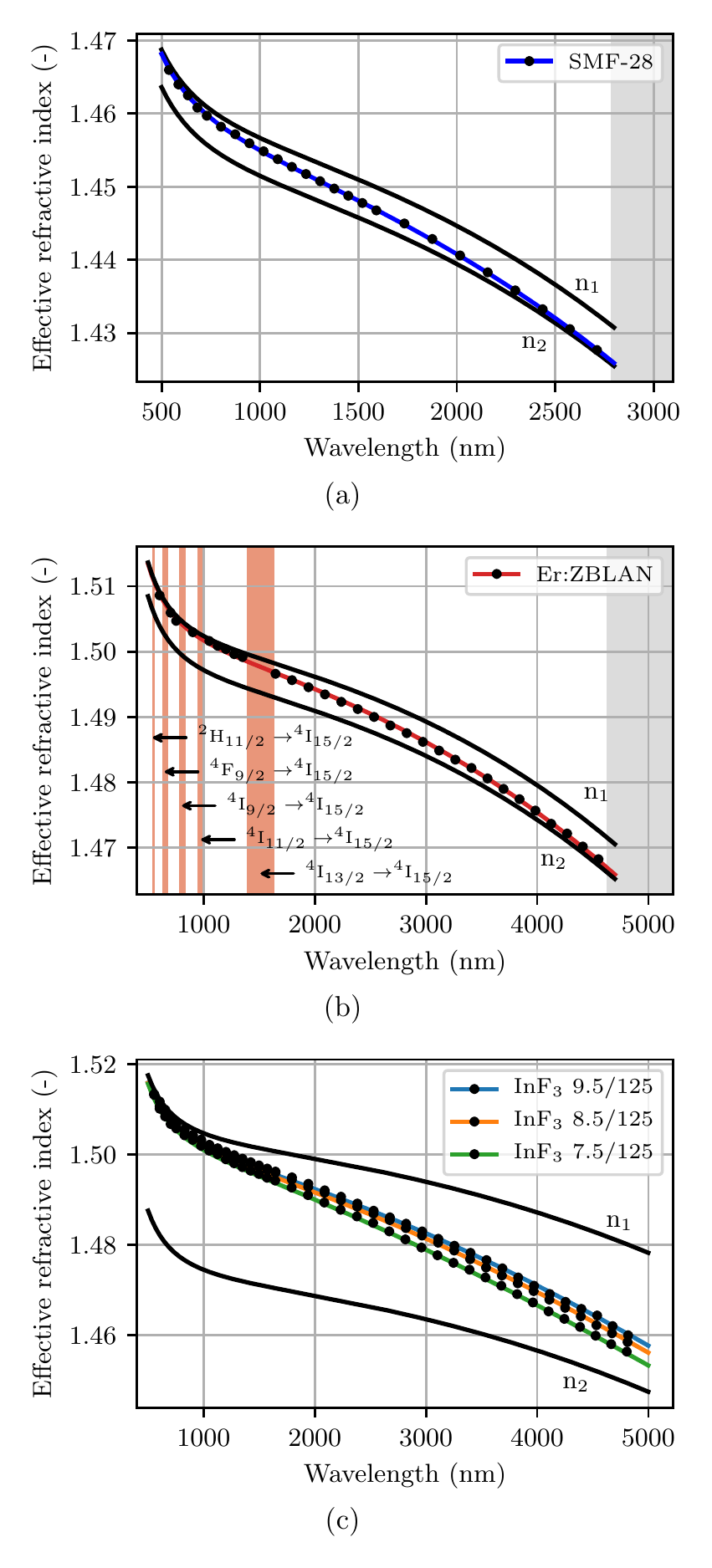}
    \caption{Measured effective refractive index of (a) the SMF-28, (b) the Er:ZBLAN fiber and (c) the three fluoroindate fibers. 
    The limit of transparency of each fibers is shown in gray, while the absorption bands of erbium are shown in red.}
    \label{f:neffsiliceZBLANInF3}
\end{figure}

The method was applied to the fibers listed in Table~\ref{t:listeFibre}. The SMF-28, a standard silica fiber made from Corning, was chosen to validate the method, in part due to its well-known parameters, its batch-to-batch consistency and for the ease of writing FBGs in silica fibers compared to fluoride fibers. Four fluoride glass fibers, all made by Le Verre Fluoré, were also chosen for this demonstration : a heavily-erbium-doped ZBLAN fiber commonly used in high power \SI{2.8}{\micro\meter} fiber lasers, and three undoped fluoroindate fibers used for SC generation in the mid-IR. FBGs were written in each fiber up to either the limit of their transparency (SMF-28 and Er:ZBLAN), or up to the last period of the phase mask (InF$_3$). Moreover, to extend the range of wavelength toward the shorter wavelengths for which the effective refractive index is to be determined, the second and third diffraction orders of the FBGs were also measured over the spectral ranges 800-\SI{1520}{\nano\meter} and 535-\SI{730}{\nano\meter}, respectively. Based on the approach detailed above, the effective refractive indices of each fiber were calculated from the measured Bragg wavelengths and are presented as black dots in Fig.~\ref{f:neffsiliceZBLANInF3}(a)-(c).

For the SMF-28 fiber, the standard 3-term Sellmeier equation for germanosilicate glass was initially used to estimate the refractive index of the core  ($n_1$) and the cladding ($n_2$), which were then used to calculate the expected effective refractive index for the LP$_{01}$ mode with good agreement~\cite{Buck_1995}. However, for the fluoride glass fibers, it was necessary to adjust the coefficients of a two-term Sellmeier equation of the form

\begin{equation}
    n^2 - 1 = \sum_{i=1}^2 \frac{A_i \lambda^2}{\lambda^2 - \lambda_i^2}\label{eq:Sellmeier}
\end{equation}

since none of the available data found in the literature would correspond to our fibers~\cite{Zhang1994, Salem2015}. In fact, the precise value of the refractive index of a fiber depends on several factors including its level of dopants, its geometry and its drawing conditions, the latter alone being able to account for a refractive index decrease as large as \SI{8e-3}{} in fluoride fibers, compare to the corresponding preform material~\cite{Nakai1990}. For this reason, for each host material, including the silica fiber used to validate the procedure, the refractive index of the cladding was adjusted with a two-term Sellmeier equation and the refractive index of the core was determined from the fiber's numerical aperture, such that the calculated $n_\eff$ matches with the measured one. The inferred Sellmeier coefficients for each host material are listed in Table~\ref{t:sellmeier}.

\begin{table}[htbp]
\centering
\caption{\bf Sellmeier coefficients for the cladding glasses of the three host materials.}
\begin{tabular}{ccccc}
\hline
Host material & $A_1$ (-) & $A_2$ (-) & $\lambda_1$ (\SI{}{\micro\meter}) & $\lambda_2$ (\SI{}{\micro\meter}) \\
\hline
SiO$_2$ & 1.1064 & 0.7250 & 0.091 & 9.11 \\
ZrF$_4$ & 1.2350 & 0.7520 & 0.090 & 14.47 \\
InF$_3$ & 1.1650 & 1.1570 & 0.100 & 20.90 \\
\hline
\end{tabular}
\label{t:sellmeier}
\end{table}

Interestingly, the asymptotical behavior of the effective refractive index at short and long wavelengths for each fiber is consistent with what one would expect from the mode confinement. At short wavelengths, the mode is expected to be tightly confined in the core, such that $n_\eff \rightarrow n_1$, while at long wavelengths, the mode is expected to be loosely confined, such that $n_\eff \rightarrow n_2$. From our measurements, the estimated cutoff wavelength of the fibers, $\lambda_C$, are \SI{1315}{\nano\meter} for the SMF-28, \SI{2.45}{\micro\meter} for the Er:ZBLAN fiber and 2.96, 3.44 and \SI{3.72}{\micro\meter} for the 7.5/125, 8.5/125 and 9.5/125 fluoroindate fibers, respectively, which are in excellent agreement with the expected values (see Table~\ref{t:listeFibre}). Below the cutoff wavelength, other distinct peaks started appearing, corresponding to higher order modes, but they were not considered in the present report.

The dispersion of the fiber, $D$, can be numerically derived from the effective refractive index~\cite{Buck_1995}:

\begin{equation}
    D = \frac{-\lambda}{c}\frac{d^2 n_\eff}{d \lambda^2}
\end{equation}

\begin{figure}[htbp]
    \centering
    \includegraphics{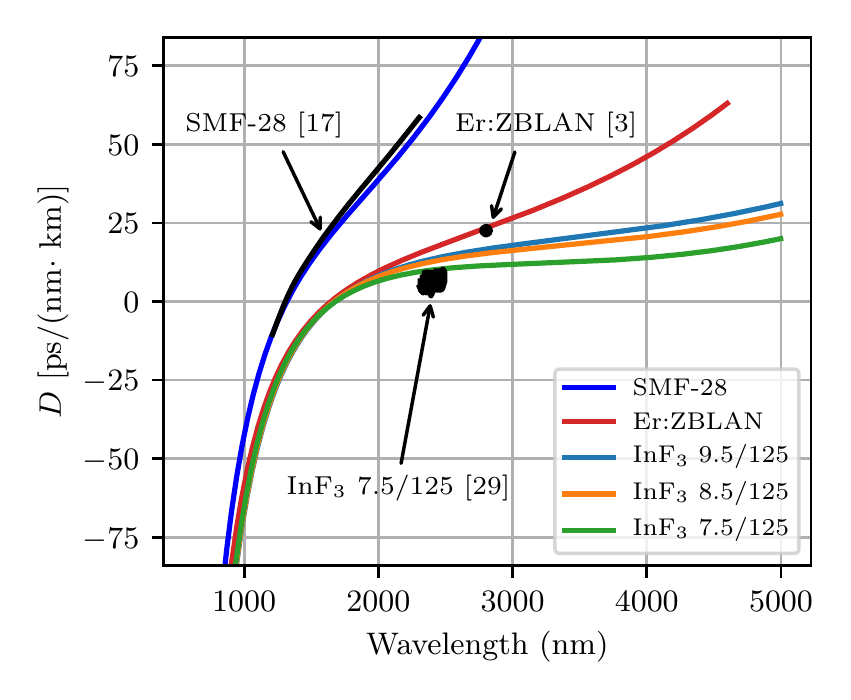}
    \caption{Calculated dispersion, $D$, from the measured effective refractive indices of Fig.~\ref{f:neffsiliceZBLANInF3}.}
    \label{f:beta2siliceZBLANInF3}
\end{figure}

Dispersion is thus simply obtained via a second order numerical derivation of the $n_\eff$ curve of Fig.~\ref{f:neffsiliceZBLANInF3}. The results are shown in Fig.~\ref{f:beta2siliceZBLANInF3}, where the dispersion of each fiber is determined up to the last measured FBG. While the SMF-28's manufacturer gives a zero-dispersion wavelength (ZDW) ranging between \SI{1302}{\nano\meter} and \SI{1322}{\nano\meter}, our measurements give a ZDW of \SI{1317.5}{\nano\meter} and the curve is in exceptional agreement with the manufacturer curve in the range 1200-\SI{1600}{\nano\meter}, while still being in good agreement up to \SI{2300}{\nano\meter}, as measured in~\cite{Ciacka2018}. Regarding the fluoride glass fibers, their dispersion is not so well known, making it hard to make a valid comparison. Nonetheless, the Er:ZBLAN fiber and the InF$_3$ fibers exhibit ZDWs around \SI{1646}{\nano\meter} and \SI{1674}{\nano\meter}, respectively, dominated by the material dispersion of the fibers, which is close to the values of \SI{1.65}{\micro\meter} and \SI{1.71}{\micro\meter} that are reported for similar fibers~\cite{Gauthier2018, Tang2016}. Moreover, measurements of the dispersion of the Er:ZBLAN fiber around \SI{2.8}{\micro\meter} and a similar 7.5/125 InF$_3$ fiber are close to our determined dispersion~\cite{Duval2015, Theberge2018}.

The dispersion inferred here is limited by a few factors. First, the refractive indices are approximated by a Sellmeier equation with two terms, i.e. one in the UV and one in the mid-IR. In reality, fluoride fibers generally have as much as five or six different glass constituents, leading to multiple resonances in the UV and the far mid-IR. Moreover, the core has a different composition than the cladding in order to increase its refractive index, leading to refractive indices that can have slightly different shapes, and a numerical aperture that slightly varies along the broad spectrum. The overall contribution of these multiple resonances was not accounted for here, but might have a small impact on the shape of the dispersion near the edges of the different curves, i.e. in the visible and the mid-IR.

It is also interesting to note that, even if in this experiment only the Bragg wavelengths of the LP$_{01}$ mode in the fibers were measured, other resonances were observed whenever the fibers were multimode. In the future, the Bragg wavelengths associated with these higher order modes could be used to increase the accuracy of the fitted refractive index for the core and the cladding by forcing the calculated effective refractive index for each mode to pass through these values too. In fact, with these values and a better post-processing of the data, we believe that all of the fiber parameters, such as its core and cladding refractive indices, its numerical aperture, its core diameter, its cutoff wavelength, etc., could be inferred with high precision.

In summary, this Letter has reported a method to accurately determine the effective refractive index and the dispersion of various optical fibers over a large bandwidth by writing arrays of FBGs using femtosecond pulses. This method was applied to silica- and fluoride-based optical fibers over almost their entire transmission range and has the potential to be extended to characterize the dispersion of other exotic fibers such as those based on tellurite and chalcogenide glass, and to precisely and fully characterize other optical and physical parameters of such fibers, such as their cutoff wavelength and their numerical aperture by measuring the Bragg wavelength associated with higher order modes propagating in the fiber.

\begin{backmatter}

\bmsection{Funding} Canada Foundation for Innovation (5180); Fonds de recherche du Québec - Nature et technologies (CO25665); Natural Sciences and Engineering Research Council of Canada (CRDPJ-543631-19, IRCPJ469414-18, RGPIN-2016-05877).

\bmsection{Acknowledgments} The authors thank Jean-Christophe Gauthier, Pascal Paradis and Vincent Fortin for helpful discussions, as well as Le Verre Fluoré for providing the InF$_3$ fiber samples.




\end{backmatter}

\bibliography{ref}

\bibliographyfullrefs{ref}


\ifthenelse{\equal{\journalref}{aop}}{%
\section*{Author Biographies}
\begingroup
\setlength\intextsep{0pt}
\begin{minipage}[t][6.3cm][t]{1.0\textwidth} 
  \begin{wrapfigure}{L}{0.25\textwidth}
    \includegraphics[width=0.25\textwidth]{john_smith.eps}
  \end{wrapfigure}
  \noindent
  {\bfseries John Smith} received his BSc (Mathematics) in 2000 from The University of Maryland. His research interests include lasers and optics.
\end{minipage}
\begin{minipage}{1.0\textwidth}
  \begin{wrapfigure}{L}{0.25\textwidth}
    \includegraphics[width=0.25\textwidth]{alice_smith.eps}
  \end{wrapfigure}
  \noindent
  {\bfseries Alice Smith} also received her BSc (Mathematics) in 2000 from The University of Maryland. Her research interests also include lasers and optics.
\end{minipage}
\endgroup
}{}

\end{document}